# Single-Shot Single-Beam Coherent Raman Scattering Thermometry Based on Air Lasing

Xu Lu[1,2], Yewei Chen[1,3], Francesco Mazza[4], Siyi He[1,2], Zihan Li[1,3], Shunlin Huang[1], Quanjun Wang[1], Ning Zhang[1,2], Bo Shen[1,5], Yuzhu Wu[1,6], Jinping Yao[1], Ya Cheng[1]

Correspondence: Jinping Yao (jinpingmrg@163.com)

[1]State Key Laboratory of High Field Laser Physics, Shanghai Institute of Optics and Fine Mechanics, Chinese Academy of Sciences, Shanghai 201800, China

[2]Center of Materials Science and Optoelectronics Engineering, University of Chinese Academy of Sciences, Beijing 100049, China

[3]School of Optical-Electrical and Computer Engineering, University of Shanghai for Science and Technology, Shanghai 200093, China

[4]Faculty of Aerospace Engineering, Delft University of Technology, Kluyverweg 1, 2629 HS Delft, The Netherlands

[5]Institute of Modern Optics, Nankai University, Tianjin 300350, China

[6]School of Microelectronics, Shanghai University, Shanghai 200444, China

**Abstract**

Thermometric techniques with high accuracy, fast response speed and ease of implementation are desirable for the study of dynamic combustion environments, transient reacting flows, and non-equilibrium plasmas. Herein, single-shot single-beam coherent Raman scattering (SS-CRS) thermometry is developed, for the first time to our knowledge, by using air lasing as a probe. It's proved that the air-lasing-assisted CRS signal has a high signal-to-noise ratio enabling single-shot measurements at a 1 kHz repetition rate. The SS-CRS thermometry consistently exhibits precision better than 2% at different temperatures, but the inaccuracy grows with the increase in temperature. The high detection precision, 1 kHz acquisition rate and easy-to-implement single-beam scheme are achieved thanks to the unique temporal, spectral and spatial characteristics of air lasing. This work opens a novel avenue for high-speed CRS thermometry, holding tremendous potential for fast diagnostics of transient reacting flows and plasmas.

**Introduction**

Determination of temperature and species concentration is of vital importance to unravel the underlying physical and chemical processes in reacting flows and plasmas, which are ubiquitous in defense, energy, space, and various other research fields[1-3]. The precise measurement of these key parameters stands as an indispensable task for the efficient and clean utilization of fuels, the design and optimization of aero-engines, etc. However, combustion diagnostics is often implemented in complex or even extreme scenarios, wherein rapid variations of temperature and combustion products pose substantial challenges to metrology. Coherent Raman scattering (CRS) spectroscopy has been proved to be a powerful non-intrusive tool in thermometry[4-9] and species detection[7-11], as it provides a strong, well-collimated signal as compared to spontaneous Raman scattering[12]. Particularly, hybrid femtosecond/picosecond (fs/ps) CRS technique, wherein the Raman coherence is interrogated by a ps pulse, has been widely applied in combustion diagnostics[3]. The narrowband probe allows us to directly record vibrational or rotational fingerprints of Raman-active molecules in the spectral domain, providing the possibility of single-shot measurements at kilohertz, or even higher, rates[13].

Nonetheless, most implementations of CRS spectroscopy necessitate two or more laser beams, where fine spatio-temporal control is essential to meet the phase-matching condition and suppress the non-resonant background[3]. In high-pressure or high-temperature turbulent environments, the CRS signal would significantly decay or even disappear due to phase mismatch caused by the beam jitter or refractive-index change. Single-beam CRS schemes, which have been widely studied, can surmount these obstacles[14-19]. Nevertheless, in traditional single-beam implementations, the probe is usually extracted from the fs pump pulse and their frequencies are very close, resulting in a poor signal-to-noise ratio (SNR). It is thus difficult to acquire pure rotational CRS signal, nor the vibrational CRS in single-shot acquisition mode.

Besides, the ps probe is usually obtained through the pulse shaping system based on a spatial light modulator (SLM). Aside from the complexity of apparatus, the low damage threshold of SLM makes it difficult to realize single-shot measurements at 1 kHz or higher rate. Up to date, single-shot single-beam coherent Raman scattering (SS-CRS) thermometry remains a big challenge and deserves further exploration.

Air lasing has emerged as a remarkable phenomenon in the realm of ultrafast optics[20-22], providing an attractive ps probe for CRS due to its inherent characteristics. In the time domain, air lasing typically exhibits a rapid rise and a slow decay, along with an intrinsic delay with respect to the pump pulse[23]. The asymmetric temporal envelope favors an efficient Raman scattering, while suppressing the non-resonant four-wave mixing background. Furthermore, the spectral bandwidth of air lasing can be as narrow as a few inverse centimeters[23], guaranteeing a high spectral resolution even in pure-rotational CRS. More importantly, the air lasing emission is coherently created along the plasma channel of the pump laser, it thus automatically overlaps with the pump beam. Therefore, air lasing is an ideal probe source for the single-beam CRS scheme. Over the past years, various air-lasing-based CRS spectroscopies have been developed[23-30] and applied to the high-sensitive detection of greenhouse gases[26, 27] and to isotope identification[27, 29]. Different from the aforesaid works for gas sensing, the main concerns of CRS thermometry for combustion diagnostics include high accuracy, fast response speed and ease of implementation. Thus, aside from single-beam scheme, single-shot measurement is further required, which has never been explored.

In this letter, single-beam CRS thermometry is developed based on air lasing. To the best of our knowledge, this is the first realization of 1 kHz single-shot CRS measurements with a single fs laser beam. Through SS-CRS thermometry, the temperature in the interaction region is retrieved by fitting the measured rotational CRS spectrum to the theoretical spectrum. Aside from its ease of implementation, this air-lasing-based thermometry also shows high measurement precision and fast response time. Therefore, it provides an advanced tool for combustion diagnostics in harsh scenarios and non-equilibrium systems[31].

**Basic principle and theoretical model**

**Basic principle**

SS-CRS is essentially a resonant four-wave-mixing process, wherein the molecular Raman coherence is excited by a broadband fs laser and subsequently probed by the narrowband ps air lasing. The basic principle is schematically shown in Fig. 1. The fs laser, after ionizing nitrogen molecule ($N_2$), serves as a degenerate pump and anti-Stokes pulse, which can excite coherent rotations of the target oxygen molecule ($O_2$). The air lasing around 428 nm, corresponding to the transition between the $B^2\Sigma_u^+(v'=0)$ and $X^2\Sigma_g^+(v=1)$ electronic states of the nitrogen ion ($N_2^+$), acts as a ps probe. When the $N_2^+$ lasing interacts with the coherently rotating molecules, the rotational coherent Stokes Raman scattering (RCSRS) signal is efficiently generated. The energy-level diagram for this process, and a typical RCSRS spectrum of $O_2$ are shown in Fig. 1c and d, respectively.

The different rotational Raman transitions are clearly resolved in the measured spectrum thanks to the narrow spectral bandwidth of the air lasing emission. The strength of each rotational Raman peak is closely related to the population difference between the initial and final rotational levels, while the rotational population itself depends on the temperature. Hence, we can retrieve the temperature in the interaction region from the RCSRS spectrum. Moreover, the measured spectrum shows a good SNR, which is attributed to the efficient rotational excitation provided by the fs laser and the unique temporal envelope of the $N_2^+$ lasing. As illustrated in Fig. 1b, the $N_2^+$ lasing has a delay of ~37 ps with respect to the pump pulse. The temporal separation of two pulses avoids the generation of non-resonant four-wave mixing, which may interfere with or even overwhelm the RCSRS signal. The $N_2^+$ lasing also shows a good spatial profile, as shown in inset of Fig. 1b, allowing an ideal spatial overlap with the pump beam. The combination of the fs pump pulse and the self-generated ps air lasing enables single-shot, single-beam operation of CRS thermometry.

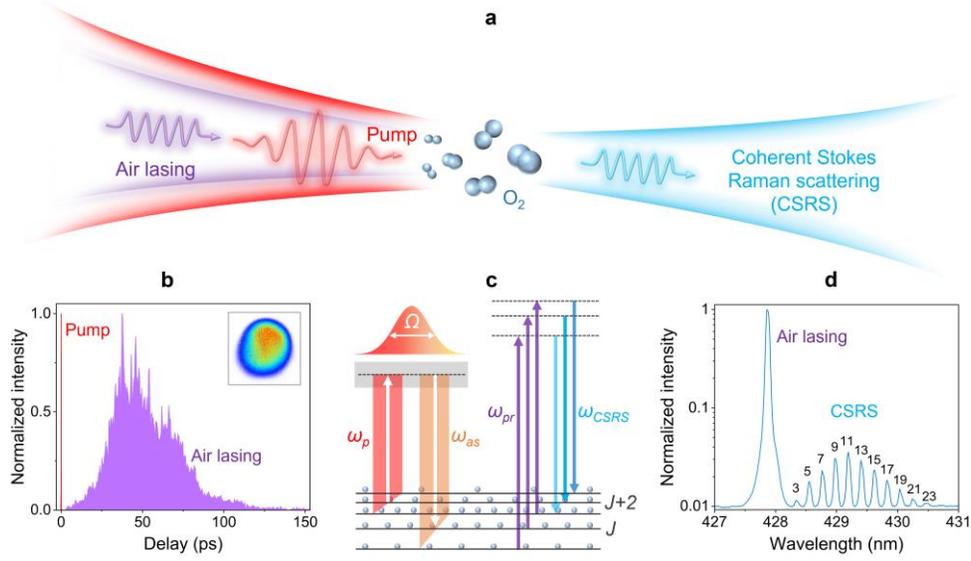

**Fig. 1 a** Schematic of basic processes of SS-CRS thermometry based on air lasing. **b** Temporal envelope of $N_2^+$ lasing and relative delay with respect to the pump pulse obtained by cross-correlation measurement. The inset shows the spatial profile of the $N_2^+$ lasing. **c** Energy-level diagram of RCSRS. **d** The typical RCSRS spectrum measured in $O_2$ at room temperature. In order to measure the air lasing and RCSRS signal simultaneously, a filter was used to attenuate the lasing signal in the data acquisition. The lower rotational state $J$ of each Raman transition is labeled on the corresponding Raman peak.

**Theoretical model**

As described above, CRS is a resonant four-wave mixing process. The corresponding third-order polarization[5, 32] can be simplified as

$$P^{(3)}(t,\tau) = \left(\frac{i}{\hbar}\right)^3 E_{pr}(t) \int_0^\infty dt' R(t') E_p^*(t+\tau-t') E_p(t+\tau-t'). \qquad (1)$$

Here, $E_p$ represents the electric field envelope of the pump pulse, which was assumed to be a Gaussian pulse with a full-width-at-half-maximum of 44 fs considering the dispersion of the window and lens. $E_{pr}$ is the electric field envelope of the probe pulse. The temporal profile of the probe pulse and the delay $\tau$ with respect to the pump were determined by the cross-correlation measurement result in Fig. 1b. The non-resonant term was neglected in Eq. 1 owing to temporal separation of the pump and probe pulses.

The molecular response function can be expressed as $R(t) = \sum I(T)e^{-i\omega t - \Gamma t}$ with transition frequency $\omega$, collision-dominated Raman linewidth $\Gamma$, and transition intensity $I(T)$[5,32]. The Raman transition frequency is defined as the energy difference between the initial and final rotational states, which were calculated using the molecular parameters accessible in the NIST website[33]. For $O_2$, each energy state is a triplet, due to the coupling of electron spin and nuclear angular momentum[34]. In the present work, the model is simplified by considering the triplets as degenerate states. Collisional effects are also ignored, i.e., $\Gamma = 0$, due to the relative low gas pressure in the furnace chamber. The transition strength $I(T)$ is related to the Raman transition cross-section $(\partial\sigma/\partial\Omega)_J$ and the population difference $\Delta\rho$ between the upper and lower states. The $\Delta\rho$ is a function of temperature $T$ based on the Boltzmann distribution. As a result, the temperature can be extracted by comparing the experimentally measured RCSRS spectrum with the theoretically calculated one. Theoretical spectra at different temperatures are obtained by $|P^{(3)}(\omega,\tau)|^2 = |\int_{-\infty}^{+\infty} P^{(3)}(t,\tau)\exp(-i\omega t)dt|^2$ at a given pump-probe delay $\tau$.

**Experimental methods**

The experiment was carried out with a commercial Ti:sapphire laser system (Legend Elite-Duo, Coherent, Inc.), which delivers ~40 fs laser pulses centered at 800 nm at 1 kHz repetition rate. The laser pulse with an energy of ~6 mJ was focused by an $f = 60$ cm lens into a gas chamber filled with $N_2$ gas at 5 mbar. In this way, $N_2^+$ lasing was coherently created along the plasma channel, giving rise to the narrowband coherent radiation around the wavelength of 428 nm. The $N_2^+$ lasing, together with the residual fs laser, were then focused by an $f$=40 cm concave mirror into a 50 mbar $O_2$ gas to produce the RCSRS signal. In order to obtain various temperatures, a tube furnace was designed to heat the $O_2$ molecules, and thermocouples were applied for the measurement and feedback control of the temperature. The temperature of the heated gas can be controlled automatically around the set target temperature with the maximum error of 5 K. The thermostatic zone was longer than 15 cm in the tube furnace, and the RCSRS signal was generated in this region. After the furnace chamber, the residual fs laser and supercontinuum emission were suppressed by a dichroic mirror, two blue glasses and a short-pass filter. The RCSRS signal of $O_2$ molecules was further filtered by a long-pass filter and detected by a spectrometer (Shamrock 500i, Andor) equipped with a 2400 grooves/mm grating and an intensified CCD (iStar, Andor). Crop mode acquisition was adopted to realize single-shot measurement at a 1 kHz rate.

**Results and Discussion**

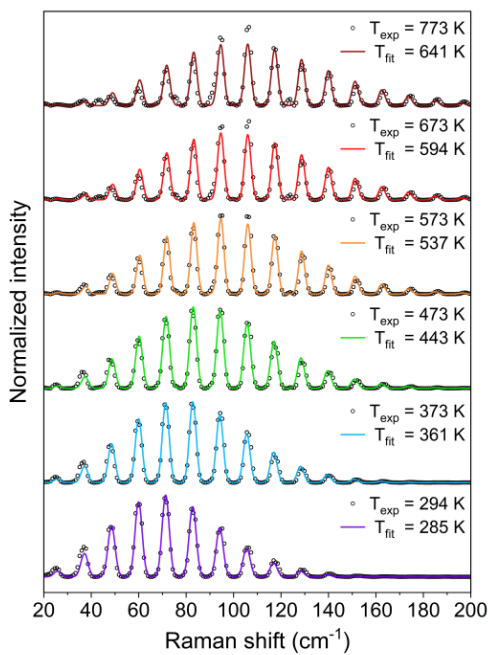

**Fig. 2** The measured RCSRS spectra at different temperatures (black circles) and the corresponding best-fit theoretical spectra (solid lines).

The typical RCSRS spectra measured at different temperatures are shown by black circles in Fig. 2. These spectra were normalized independently, and the corresponding baseline was subtracted. Each RCSRS spectrum in the figure is the average result of 1000 consecutive single-shot spectra captured within one second. We can clearly see that each RCSRS spectrum consists of a series of equally spaced peaks. Their frequency shifts with respect to the $N_2^+$ lasing are in good agreement with the approximate theoretical formula $\Delta\tilde{\lambda} = (4J + 6)B$, where $B = 1.43 \text{ cm}^{-1}$ is the rotational constant of the $O_2$ molecule[33], and $J$ is the rotational quantum number of the lower rotational state. As the temperature increases, the measured RCSRS spectrum covers a broader range and the strongest Raman line shifts towards higher rotational quantum states. These variations are attributed to the change of population distribution on various rotational states caused by the temperature. It is well known that the rotational population of molecules in thermal equilibrium follows the Boltzmann distribution. When the degeneracy of the rotational levels is also taken into account, more molecules populate higher rotational states at higher temperature. Besides, since the gas pressure of furnace chamber was kept at 50 mbar at all temperatures, the RCSRS signal gets weaker as the temperature increases due to the decrease in gas density. Here, it can be clearly seen that in spite of temperatures above 500 K, the SNRs of measured RCSRS spectra are still sufficiently high.

The experimental results in Fig. 2 qualitatively reflect the temperature variation in the interaction region. Quantitative results were obtained by fitting the measured spectra with the theoretical spectra calculated by the above model. The solid lines are the best-fit theoretical spectra, which are achieved by optimizing the temperature. The best-fit temperature is given in the corresponding panel. Clearly, the fitted spectra reproduce the main features of the measured RCSRS spectra, including the relative intensity of all Raman peaks as well as the spectral position and width of each peak. The residuals between theoretical and experimental spectra are very small expect for 1~2 Raman peaks at temperatures above 500 K.

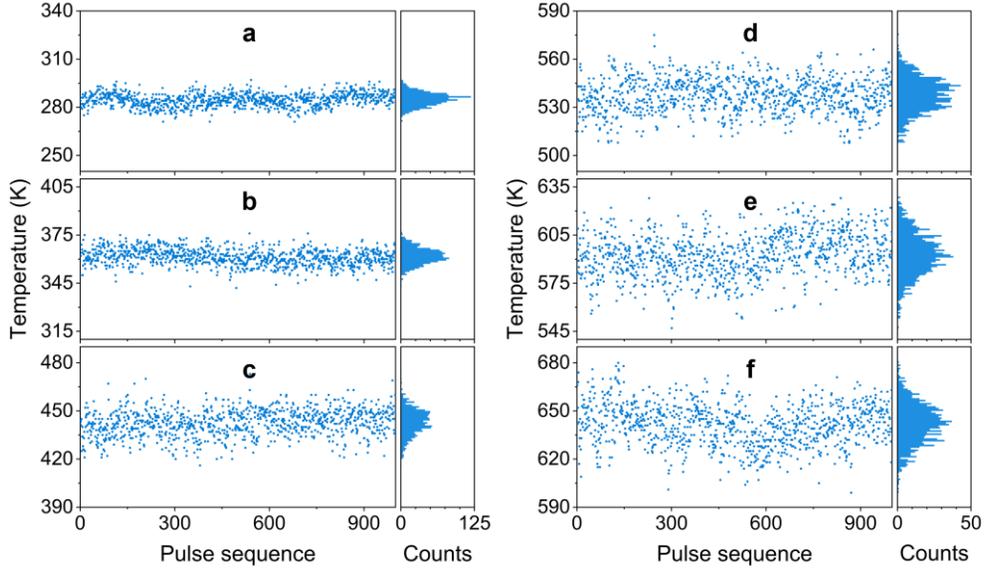

**Fig. 3** The best-fit temperatures of 1000 consecutive single-shot spectra and the corresponding histograms at different furnace temperatures: **a** 294 K, **b** 373 K, **c** 473 K, **d** 573 K, **e** 673 K and **f** 773K.

The single-shot results at different gas temperatures are shown in Fig. 3a-f. The corresponding histograms of the best-fit temperatures using 1000 consecutive single-shot spectra are illustrated in the right panels. The temperature step is taken as 1 K in the fitting process. It can be clearly seen that these histograms approximately obey the normal distribution with different mean values and standard deviations. At room temperature (294 K), the best-fit temperatures are concentrated in the vicinity of 286 K. As the target temperature grows, the best-fit temperatures of 1000 single-shot spectra show a broader distribution. To perform quantitative analysis, Fig. 4a compares the average value $T_{\text{mean}}$ of the best-fit temperatures extracted from 1000 single-shot spectra with the target temperature $T_0$. The standard deviation $T_{\text{std}}$ of the best-fit temperatures is marked by an error bar. Figure 4b shows the inaccuracy and precision of SS-CRS thermometry as a function of the target temperature. Here, the measurement inaccuracy is characterized by relative error $|T_0 - T_{\text{mean}}|/T_0$, while the precision is represented by the relative standard deviation $T_{\text{std}}/T_0$.

The quantitative results in Fig. 4a illustrate that the best-fit temperature gradually deviates from the target temperature as the latter grows. As shown in Fig. 4b, the measurement inaccuracy increases to 17.0% at 773 K from 3.2% at room temperature. The significant increase of measurement error at high temperature could be caused by multiple factors, which are discussed later. Compared with the inaccuracy, the precision always remains below 2% for all temperatures, as indicated by the black squares. Such high precision means that the shot-to-shot fluctuations of the RCSRS signal are small, which shows the obvious advantage of the SS-CRS scheme.

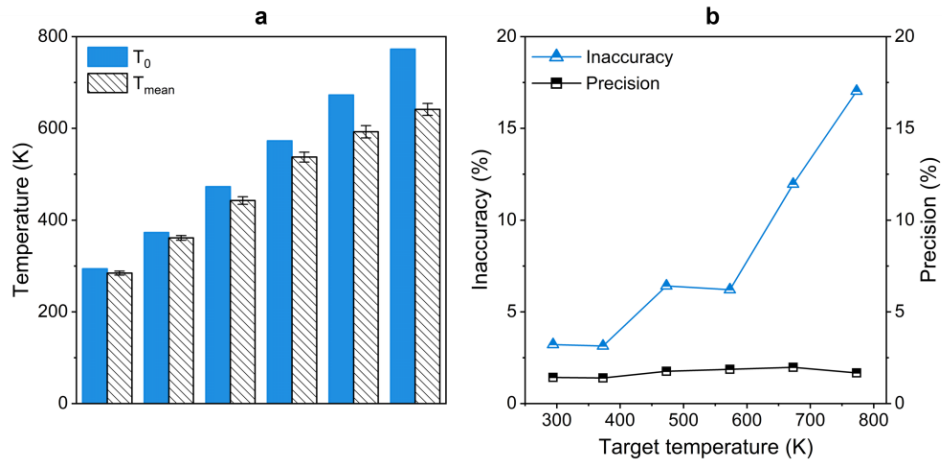

**Fig. 4 a** The comparison between the best-fit temperature extracted from single-shot RCSRS spectra and the target temperature. **b** The inaccuracy and precision of SS-CRS thermometry as a function of the target temperature.

The above results demonstrate the feasibility of 1 kHz, single-shot measurements via SS-CRS thermometry. This state-of-the-art thermometry ingeniously combines the advantages of the fs pump laser and the self-generated air lasing. Especially, the time-frequency characteristics of the latter endow this technique with the capability of single-shot, single-beam measurement. This novel scheme not only greatly simplifies the configuration of CRS thermometry, but also allows for combustion diagnostics in harsh environments or at standoff locations. It benefits from the inherent spatial overlap of the pump, Stokes/anti-Stokes and probe pulses and from the elimination of non-resonant background due to the intrinsic delay of the pump pulse and air lasing emission.

As the first proof-of-principle demonstration, both accuracy and precision of SS-CRS thermometry reach a satisfactory level at temperatures up to 100 °C[8]. Nevertheless, the accuracy degrades at higher temperatures, as shown in Fig. 4b. Several factors could be responsible for this. First, the gas pressure is kept at 50 mbar at all temperatures. The decrease of gas density with the increasing temperature deteriorates the SNR of the RCSRS signal, which could affect the fitted temperature and measurement accuracy. Second, spatio-temporal dynamics of the fs laser is not taken into account in our simulation, which could influence the Raman excitation[35,36]. Last but not least, $N_2$ molecules are inevitably excited during the air lasing generation. Contamination of the $O_2$ RCSRS spectra by of a weak $N_2$ RCSRS signal could also contribute to the error in the temperature fitting. For practical applications in high-temperature, high-pressure combustion scenarios, we need to further improve the accuracy of the temperature measurement by considering or minimizing the influences of aforementioned factors. The precision of the present technique is comparable to the results of conventional fs/ps CRS thermometry[8]. Indeed, such a single-beam CRS thermometry will demonstrate greater superiority while it is applied in the diagnostics of dynamic combustion environments or high-speed reacting flows.

In addition, the spatial resolution of a thermometric technique plays a crucial role in the diagnostics of turbulent combustion fields with temperature gradients. In the present single-beam scheme, the collinear configuration of the air lasing emission and the fs pump laser results in poor spatial resolution. Indeed, the temperature extracted from the RCSRS spectra is the integral result over the interaction region. In other words, the spatial resolution is determined by the focusing geometry. If a tighter focusing lens or a telescope system is used, the spatial resolution can be further improved. We can even obtain a spatial resolution comparable to that of a BOXCARS configuration[5] by using the spatio-temporal focusing scheme[37]. These technical improvements will help us greatly improve the spatial resolution, the response speed, the measurement accuracy and precision of SS-CRS thermometry. We believe that air-lasing-based thermometry can be applied in the studies of dynamic high-temperature combustion environments, transient reacting flows, and non-equilibrium plasmas.

**Conclusions**

In summary, we developed 1 kHz single-shot coherent Raman thermometry with a single femtosecond laser beam. This single-shot, single-beam thermometric technique fully utilizes the spatial, spectral and temporal properties of air lasing. The single-beam scheme largely simplifies the apparatus, and overcomes the difficulty of the spatio-temporal control of multiple beams in high-temperature turbulent environments. It is proven that the inaccuracy and precision of the thermometry reach 3.2% and 1.4% at room temperature, respectively, while operating in the 1 kHz, single-shot acquisition mode. As the temperature increases, the precision remains almost unchanged, whereas the accuracy gradually worsens. Although the accuracy is not satisfactory at high temperatures in the present stage, some technical improvements promise to overcome this limitation. Air-lasing-based SS-CRS thermometry will show promising applications in the investigation of reacting flows and plasmas.

**Acknowledgements**

This work is supported by National Natural Science Foundation of China (Grant No. 12034013, No. 12274428, and No. 12374320), Youth Innovation Promotion Association of Chinese Academy of Sciences (CAS) (Grant No. Y2022072), CAS Project for Young Scientists in Basic Research (Grant No. YSBR-042), Natural Science Foundation of Shanghai (Grant No. 22ZR1481600 and No. 23ZR1471700), and the Dutch Research Council (NWO) (Grant No. AES-15690).

**Author contributions**

J.Y. conceived the idea and designed the experiment. Y.C. supervised the project. X.L., Y.C., S.H. and Z.L. performed the experiment. X.L. and Y.C. analyzed the results. All authors contributed to the preparation of the manuscript and commented on the manuscript.

**Data availability**

Data underlying the results presented in this paper are not publicly available at this time but may be obtained from the authors upon reasonable request.

**Competing interests**

The authors declare no conflicts of interest.